\documentclass[onecolumn,12pt,journal,final]{IEEEtran}

\usepackage{amsfonts}
\usepackage[dvips]{graphicx}
\usepackage{times}
\usepackage{cite}
\usepackage{amsmath}
\usepackage{cases}
\usepackage{color}
\usepackage{array}
\usepackage{algorithm}
\usepackage{algorithmic}
\usepackage{dsfont}
\usepackage{amssymb}
\usepackage{stfloats}
\usepackage{graphicx}
\usepackage{footnote}
\usepackage{booktabs}
\usepackage{array}
\usepackage{bm}
\usepackage{stfloats}
\usepackage{diagbox}

\newtheorem{theorem}{Theorem}
\newtheorem{remark}{Remark}

\newtheorem{definition}{Definition}

\newtheorem{corollary}{Corollary}

\begin{document}

\title{Treating Content Delivery in Multi-Antenna Coded Caching as General Message Sets Transmission: A DoF Region Perspective}
\author{\IEEEauthorblockN{Youlong Cao, \IEEEmembership{Student Member,~IEEE} and Meixia Tao, \IEEEmembership{Fellow,~IEEE}\\}

\thanks{This work was supported by the NSF of China under grant 61571299 and by STCSM 18DZ2270700. This work was presented in part at the 2017 IEEE Global Communications Conference \cite{long_glo17}. The authors are with Shanghai Institute for Advanced Communication and Data Science, and Department of Electrical Engineering, Shanghai Jiao Tong University, Shanghai 200240, China. Emails: \{caoyoulong, mxtao\}@sjtu.edu.cn.}
}
\maketitle

\begin{abstract}
Coded caching can create coded multicasting thus significantly accelerates content delivery in broadcast channels with receiver caches. While the original delivery scheme in coded caching multicasts each coded message sequentially, it is not optimal for multiple-input multiple-output (MIMO) broadcast channels. This work aims to investigate the full spatial multiplexing gain in multi-antenna coded caching by transmitting all coded messages concurrently. In specific, we propose to treat the content delivery as the transmission problem with general message sets where all possible messages are present, each with different length and intended for different user set. We first obtain inner and outer bounds of the degrees of freedom (DoF) region of a $K$-user $(M,N)$ broadcast channel with general message sets, with $M$ and $N$ being the number of transmit and receive antennas, respectively. Then for any given set of coded messages, we find its minimum normalized delivery time (NDT) by searching the optimal DoF tuple in the DoF regions. The obtained minimum NDT is optimal at antenna configuration $\frac{M}{N} \in (0,1]\cup[K, \infty)$ and is within a multiplicative gap of $\frac{M}{N}$ to optimum at $\frac{M}{N} \in (1,K)$. Our NDT results can be evaluated for any user demand with both centralized and decentralized cache placement.
\end{abstract}


\section{Introduction}
Over the last decade, video streaming has become a prominent component in the mobile data traffic. It amounts for more than 60$\%$ of the total mobile data load in 2018 and is foreseen to contribute 74$\%$ in 2024 \cite{ericsson}. Caching popular video contents at the edge of wireless networks is an effective approach to accelerate content delivery by exploiting the asynchronous content reuse among multiple users \cite{caching1,caching2,liu_tao}. The fundamental gain of caching is first studied in an information-theoretic framework, known as coded caching \cite{Alilimits}. The original coded caching is proposed for an ideal broadcast channel where a server wishes to deliver requested contents to multiple cache-enabled users over a shared error-free link. The system operates in two phases, a cache placement phase and a content delivery phase. The former decides what segments of each file to store at each cache node in either centralized \cite{Alilimits} or decentralized \cite{Alidecentralized} manners, prior to user demands; the latter determines how to generate and transmit codewords upon user requests. The works \cite{Alilimits}, \cite{Alidecentralized} show that coded caching can create coded multicasting for content delivery even when user demands are different and thus achieve global caching gain. Later, coded caching has been extended in various wireless network topologies, such as combination networks \cite{combination_network1,combination_network2}, wireless interference networks \cite{xu,xu2,Ali3}, partially connected networks \cite{xu_glo17,Xinping_topo}, device-to-device networks \cite{d2d_1,d2d_twc1,d2d_twc2}, and fog radio access networks (F-RANs) \cite{fogRan,fogran2,jingjingMIMOIC}.

Among the various extensions of coded caching, one important case is multi-antenna coded caching \cite{yangsheng2,Multi_server_coded_caching,Caire2017multi,multiantennajournal,MISO_Petros_Elia,long_journal,piovano2017generalized,MISO_overloaded_BC}, where the interplay of caching and multiple-input multiple-output (MIMO) is studied. It is well known that the multicast rate of a fading broadcast channel is limited by the channel quality of the weakest user. The performance of coded caching can thus degenerate when the number of users grows. The work \cite{yangsheng2} shows that by employing multiple antennas at the transmitter, a nonvanishing multicast rate can be obtained, yielding a scalable coded caching scheme. In \cite{Multi_server_coded_caching,Caire2017multi,multiantennajournal,MISO_Petros_Elia}, the authors utilize multiple antennas to multiplex multiple coded messages by using zero-forcing and show that higher transmission rate can be achieved over single-group multicast. As an extension of coded caching in interference networks where all the transmitters and receivers have caches, the work \cite{long_journal} studies the caching gain when employing multiple antennas at all nodes. It shows that the content delivery time of the cache-aided MIMO interference network is piecewise-linearly decreasing with the cache size and inversely proportional to the number of antennas. The work \cite{piovano2017generalized,MISO_overloaded_BC} exploits the coded caching gain and spatial multiplexing gain induced by multiple antennas with partial knowledge of channel state information (CSI) and various channel strength levels.

 However, the aforementioned achievable schemes for multi-antenna coded caching do not fully exploit the spatial multiplexing gain. In specific, the schemes in \cite{Multi_server_coded_caching,Caire2017multi,multiantennajournal,MISO_Petros_Elia,long_journal} transmit the coded messages sequentially group by group, each having the same message length and intended for multiple symmetric user sets with the same size. In decentralized caching, due to that each user caches file bits independently at random, it is very likely to generate coded messages in the delivery phase with arbitrary lengths and intended for any possible user sets. Thus, these group-by-group transmission schemes do not exploit the spatial multiplexing gain among asymmetric user sets and can be far from optimal in a decentralized caching system. The works in \cite{yangsheng2,piovano2017generalized,MISO_overloaded_BC} adopt a hybrid multicast/unicast transmission scheme where one coded (multicast) message is transmitted together with multiple uncoded (unicast) messages. This hybrid transmission scheme is heuristic and cannot ensure full multiplexing gain.

The aim of this work is to investigate a transmission scheme that can fully utilize spatial multiplexing gain in multi-antenna coded caching, especially with decentralized cache placement. Towards this end, we propose to treat the delivery of coded messages, which are intended for different user sets and have different lengths, as transmission with general message sets. We consider a $K$-user $(M,N)$ cache-aided MIMO broadcast channel where a base station (BS) equipped with $M$ antennas communicates with $K$ cache-enabled users, each equipped with $N$ antennas. We would remark that there are some related but different efforts on studying the capacity of communication channels with general message sets. Some earlier works have investigated the capacity of several broadcast channels with general message sets, including the $K$-user SISO channel \cite{alphafair_coded_caching} and the two-user MIMO channel \cite{two_user_bc_gms}. The work \cite{compound_bc} characterizes the degrees of freedom (DoF) region of the $K$-user compound MISO broadcast channel, which can be regarded as a broadcast channel with multiple multicast groups but without intersection among different multicast groups. The authors in \cite{lekegeneralmessage} consider a MIMO interference channel with general message sets where each transmitter emits an independent message and each receiver requests an arbitrary subset of the messages. The work \cite{zhengdaogeneralmessage} investigates the general message sets problem in a single-antenna X channel.  The main distinction of this paper from the previous works \cite{alphafair_coded_caching,two_user_bc_gms,compound_bc,lekegeneralmessage,zhengdaogeneralmessage} lies at the channel model. We consider the $K$-user MIMO broadcast channel with general message sets, whose DoF region, to our best knowledge, has not been studied before.

Note that there are plenty of coded caching works, e.g.,  \cite{yangsheng2,Multi_server_coded_caching,Caire2017multi,multiantennajournal,piovano2017generalized,MISO_overloaded_BC}, adopting a separation approach of caching and delivery. Namely, they only focus on the delivery of coded messages by using the same cache placement and coded message generation as the ones proposed in \cite{Alilimits,Alidecentralized}. In the recent work \cite{seperation_priciple}, this separation approach is shown to be order-optimal for many network topologies. In wireless communication scenario, it is more practical to consider such a separation principle since prefetching usually happens in a large time scale while the content delivery should be optimized in a small time scale according the specific channel information. Motivated by this consideration, in this work, we follow this separation principle and focus on the efficient delivery of a given set of coded messages in multi-antenna coded caching.

 The main contributions of this paper are summarized as follows:

 $\bullet$ \emph{DoF region with general message sets}: We obtain both an inner bound and an outer bound of DoF region for the MIMO broadcast channel with general message sets at any $K,M$ and $N$. The proposed achievable scheme is based on linear transmit precoding and receive combining. For the antenna configuration $\frac{M}{N}\in\left(0,1\right]$, each user uses a zero-forcing based receive combining matrix to decode the desired messages. For the antenna configuration $\frac{M}{N}\in\left(1,K\right]$, the main idea is to equalize the effective channels of all the users within each multicast group by designing their receive combining matrices and then use a zero-forcing based transmit precoding matrix at the BS to cancel the interference among all the multicast groups. For the antenna configuration $\frac{M}{N}\in\left(K,\infty\right)$, we concatenate the signals to be decoded by each user as a meta-signal and then regard the broadcast channel with general message sets as the conventional broadcast channel with private message only. We show that for antenna configuration $\frac{M}{N} \in (0,1]\cup[K, \infty)$, the inner and outer bounds coincide and hence are the optimal DoF region. In the special case with two users $K=2$, the obtained DoF region is globally optimal at all $M$ and $N$. Note that this DoF region analysis is motivated from multi-antenna coded caching, but the results are for a general MIMO broadcast channel with general message sets.

 $\bullet$ \emph{Normalized delivery time (NDT) at any given user demand}: By using the obtained inner and outer bounds of DoF region with general message sets, we obtain both upper and lower bounds of the minimum NDT in the considered cache-aided MIMO broadcast channel. Unlike the previous NDT analysis that aims to minimize the  delivery time for the worst-case user demand, our NDT analysis is to minimize the delivery time for any given user demand at any given realization of cache placement. We show that the upper and lower bounds of NDT are the same and thus optimal at antenna configuration $\frac{M}{N} \in (0,1]\cup[K, \infty)$ and they are within a multiplicative gap of $\frac{M}{N}$ when $\frac{M}{N}\in(1,K)$. We also obtain closed-form NDT expressions for some special cases.

The remainder of the paper is organized as follows. In Section II, we introduce the system model of the cache-aided MIMO broadcast channel. Section III presents the DoF region of MIMO broadcast channel with general message sets. The achievable minimum NDT results and optimality analysis are given in Section IV and V, respectively.  Section VI shows some numerical results. Section VII concludes this paper.

Notations: $x$, ${\bf x}$, ${\bf X}$ and $\cal{X}$ denote scalar, vector, matrix and set (region), respectively. $(\cdot)^{T}$ and $(\cdot)^{\dag}$  denote the transpose and the Moore-Penrose pseudo-inverse, respectively. $\text{span}({\bf X})$ and ${\text{null} ({\bf X})}$ stand for the column span space and the null space of the matrix ${\bf X}$, respectively. $\text{rank}({\bf X})$ denotes the rank of the matrix ${\bf X}$. conv$(\cal{X})$ denotes the convex hull of the region $\cal{X}$. ${\bf x}./{\bf y}$ denotes the element-by-element division. $\parallel \bf x \parallel_{\infty}$ denotes the infinity norm of the vector $\bf x$. $[n]$ denotes the set $\{1,2,\cdots,n\}$ where $n$ is an integer. $\binom{n}{m}=\frac{n!}{(n-m)!m!}$ denotes the binomial coefficient indexed by integers $n$ and $m$. ${\bf 1}(\cdot)$ denotes the indicator function.

\section{Problem Setting}
   \begin{figure}[t]
    \begin{centering}
  \includegraphics[scale=0.42]{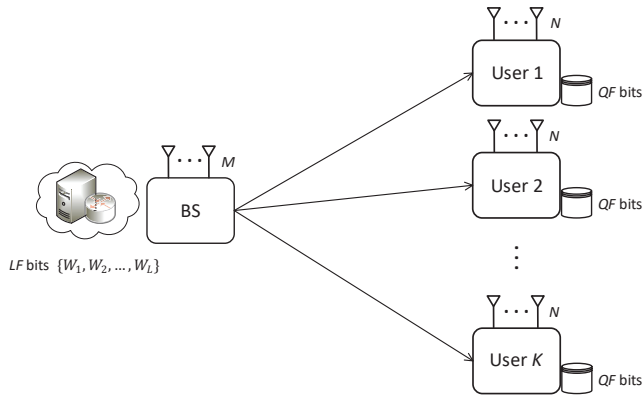}
   \caption{The $K$-user $(M,N)$ cache-aided MIMO broadcast channel.} \label{model}
   \end{centering}
    \end{figure}
Consider a $K$-user $(M,N)$ cache-aided MIMO broadcast channel as shown in Fig. \ref{model}, where a BS equipped with $M$ antennas communicates with $K$ users, each equipped with $N$ antennas, over a shared wireless link. The BS has access to a library of $L$ $(L\geq K)$ files, denoted by $\{W_1,W_2,\ldots,W_L\}$, each of length $F$ bits. Each user is equipped with a local cache and can cache up to $QF$ bits, where $Q < L $. The normalized cache size is defined as $\mu\triangleq\frac{Q}{L}$, which represents the fraction of the library that each user can store locally.

The system operates in two phases, a \emph{cache placement phase} and a \emph{content delivery phase}. In the cache placement phase, each user independently chooses $\mu F$ bits uniformly at random from each file to cache without knowing the future user demands. Let $W_{\ell, \cal K}$ with $\ell\in[L]$ denote the subfile of $W_{\ell}$ cached exclusively in user set $\mathcal K\subseteq[K]$. The length of $W_{\ell, \mathcal K}$ is random, depending on the specific realization of cache placement. Note that in many existing works, e.g., \cite{Alidecentralized}, the file size $F$ is assumed to be large enough so that the length of each $W_{\ell, \mathcal K}$ can be approximated as a constant for analytical tractability by the law of large numbers. However, in this paper, we do not need this assumption. The cached content of user $i\in[K]$ after decentralized cache placement is given by
\begin{align}
  U_i = \left\{W_{\ell, \mathcal K}:  \forall {\mathcal K}\subseteq [K], \forall \mathcal K \ni i, \forall \ell\in [L] \right\}.
\end{align}
In the content delivery phase, each user $i$ requests one file $W_{r_{i}}$ with $r_i\in[L]$ from the library. We define ${\bf r}=(r_1,r_2,\ldots,r_K)^T$ as the vector of all user demands. Upon receiving the user demands $\bf r$, the BS generates a set of coded messages by utilizing the user side information established in the cache placement phase. As a rule of thumb, the set of coded messages is generated by  \cite{Alidecentralized}:
 \begin{equation} \label{generate_coded_message}
   W_{\mathcal A}^{\oplus}=\oplus_{i\in{\mathcal A}} W_{r_i,  {\mathcal A}\backslash\{i\}}.
 \end{equation}
where $\oplus$ denotes the bit-wise XOR operation, and $W_{\mathcal A}^{\oplus}$ is intended to user set ${\mathcal A} \subset [K]$. The above coded message is a combining of the subfiles $\{W_{r_i, {\mathcal A}\backslash\{i\}}\}$, each desired by user $i$ and cached at all other users $i'\in {\mathcal A}$, $i'\neq i$, for all $i\in \mathcal A$. In general, finding the optimal XOR combining to minimize the total traffic load is challenging due to the randomness of the length of each $W_{r_i, {\mathcal A}\backslash\{i\}}$. Several modified schemes have been proposed in \cite{long_glo17,Mingyuecodedcaching3,yu2016exact,naifu}. Following the separation approach of caching and delivery, the current work focuses on the physical layer transmission of these coded messages $\{W_{\mathcal A}^{\oplus}\}$ and is suitable for any coding scheme under any specific realization of decentralized cache placement and with any file length. We consider the most general case where all the $2^K-1$ possible coded messages $\{W_{\mathcal A}^{\oplus}\}$ are present, each with length $a_{\mathcal A}F$ bits. Here $a_{\mathcal A}$ is a fraction and depends on the specific realization of the decentralized cache placement, user demand vector $\bf r$, and the XOR combining scheme. Each user $i$ needs to decode all coded messages $\{W_{\mathcal A}^{\oplus}\}$ for ${\mathcal A}\ni i$.

In the physical layer, the BS applies coding and modulation on $\{W_{\mathcal A}^{\oplus}\}$ and transmits them over $T$ time slots. The input-output relationship of the broadcast channel at each time slot $t\in[T]$ in the delivery phase is given by
\begin{equation}
{\bf y}_{i}(t)={\bf H}_{i}{\bf x}(t)+{\bf n}_{i}(t),\quad \forall i \in [K].
\end{equation}
Here ${\bf y}_{i}(t)\in{\mathbb C}^{N\times 1}$ is the received signal of user $i$; ${\bf H}_{i}\in{\mathbb C}^{N\times M}$ is the channel matrix from the BS to user $i$, whose entries are drawn independently and identically distributed (i.i.d) from a continuous distribution, and remain invariant within each codeword transmission;  ${\bf n}_{i}(t)\in{\mathbb C}^{N\times 1}$ denotes the additive white Gaussian noise (AWGN) vector at user $i$, with each element being independent and having zero mean and unit variance; the transmitted signal ${\bf x}(t)\in{\mathbb C}^{M\times 1}$ is subject to a power constraint
\begin{equation}
  \text{tr}\left[{\bf x}(t){\bf x}^H(t)\right]\leq P.
\end{equation}
 Each user $i$ recovers its requested file from the received signal along with its own local cache content $U_i$. We assume that the perfect CSI is available at the BS and all the users.

The system error probability is defined as
\begin{equation}
P_e=\max\limits_{{\bf r}\in[L]^K}\max\limits_{i\in[K]} \mathbb{P}({\hat W}_{i}\neq W_{r_i}),
\end{equation}
where ${\hat W}_{i}$ is the estimated file of user $i$. The delivery time $T$ is achievable if, for almost all channel realizations, the error probability $P_e$ approaches 0 when $F\rightarrow \infty$.

In this paper, we use the \emph{normalized delivery time} (NDT) introduced in \cite{simeone} to characterize the asymptotic performance of cache-aided wireless networks in the high signal-to-noise ratio (SNR) region and large file size.
\begin{definition}
   Let $T(\bf f)$ denote an achievable delivery time for a given set of coded messages $\{W^{\oplus}_{\mathcal A}\}$, with message length vector ${\bf f}=\left[a_{{\mathcal A}_1},a_{{\mathcal A}_2},a_{{\mathcal A}_3},\cdots,a_{{\mathcal A}_{2^K-1}} \right]^T$. The corresponding NDT is defined as
\begin{equation} \label{eqn NDTtau}
\tau({\bf f})\triangleq\lim\limits_{P \to \infty}\lim\limits_{F \to \infty}\sup\frac{T(\bf f)}{F/\log P}.
\end{equation}
\end{definition}

Note that the original NDT definition in \cite{simeone} is for the worst-case user demand while the above definition is for a given user demand and the NDT is a function of the coded message length vector $\bf f$ associated with the given user demand. This is because while previous works focus on the joint design of cache placement and content delivery to minimize the worst-case delivery time, we adopt the separation approach of caching and delivery in \cite{seperation_priciple} and focus on the content delivery to minimize the delivery time for any given user demand at any given realization of cache placement. By minimizing $\tau(\bf f)$ for each $\bf f$, we are then able to optimize the overall latency performance of the considered system at all possible user demands.

\setcounter{equation}{7}
\begin{figure*}[!t]
  \begin{align}\label{defintion_of_DoF_region}
{\mathcal D}=\bigg\{{\bf d}\in {\mathbb R}^{2^K-1}_{+}:\forall \,(w_{{\mathcal A}_1},w_{{\mathcal A}_2},&\cdots,w_{{\mathcal A}_{2^K-1}})\in {\mathbb R}^{2^K-1}_{+}, \nonumber \\
  &\sum\limits_{{\mathcal A}:{\mathcal A}\in\Psi} w_{{\mathcal A}}d_{{\mathcal A}}\leq \mathop{\lim \sup}\limits_{P\rightarrow \infty} \left[\sup_{{\mathcal R}(P)\in{\mathcal C}(P)}\left[\sum\limits_{{\mathcal A}:{\mathcal A}\in\Psi}w_{{\mathcal A}}R_{{\mathcal A}}(P)\right]\frac{1}{\log P}\right] \bigg\}.
\end{align}
\hrulefill
\end{figure*}

\section{DoF Region with General Message Sets}
It becomes apparent from the previous section that transmitting the set of coded messages $\{W_{\mathcal A}^{\oplus}\}$, each with length $a_{\mathcal A}F$ bits and intended for user set ${\mathcal A}\subseteq[K]$ has turned the original channel into a channel with general message sets. To characterize the minimum delivery time of these coded messages, it is essential to find the capacity region of the $K$-user $(M,N)$ MIMO broadcast channel with general message sets. In this work, we focus on the analysis of the capacity region in the high SNR scenario, i.e., DoF region, for analytical tractability. Both an outer bound and an inner bound of the DoF region will be given in this section.

Let $\Psi =\{{\mathcal A}_1, {\mathcal A}_2, \cdots, {\mathcal A}_{2^K-1}\}$ denote the set of all the $2^K-1$ user sets. Throughout this paper, each user set is also called a multicast group. Let $R_{\mathcal A}$ denote the transmission rate of the message $W_{\mathcal A}^{\oplus}$ intended to user set $\mathcal A\in \Psi$. The capacity region ${\mathcal C}(P)$ is the set of all achievable rate tuples ${\mathcal R}(P)=\big(R_{{\mathcal A}_1}(P),R_{{\mathcal A}_2}(P),\cdots,$ $R_{{\mathcal A}_{2^K-1}}(P)\big)$.  We define the DoF tuple as
\setcounter{equation}{6}
\begin{align}
  {\bf d}=\left(d_{{\mathcal A}_1},d_{{\mathcal A}_2},\cdots,d_{{\mathcal A}_{2^K-1}}\right).
\end{align}
Define the corresponding DoF region as shown in \eqref{defintion_of_DoF_region} at the top of this page.
\setcounter{equation}{8}

\subsection{Main Results}
\begin{theorem}[Outer Bound]\label{theorem1}
  An outer bound of the DoF region of the $K$-user $(M,N)$ MIMO broadcast channel with general message sets, denoted as ${\mathcal D}_{\text{out}}$ is given by
   \begin{subequations}
\begin{align}
{\mathcal D}_{\text{out}}=\Big\{&{\bf d } \in {\mathbb R}^{2^K-1}_{+}: \nonumber \\
&\sum\limits_{{\mathcal A}:i\in{\mathcal A}\in\Psi} d_{\mathcal{A}}\leq N,\quad  \forall i\in[K]  \label{cut_set_user}\\
&\sum\limits_{{\mathcal A}:{\mathcal A}\in \Psi} d_{\mathcal A}\leq M  \label{cut_set_BS}\Big\}.
\end{align}
\end{subequations}
\end{theorem}

 The above outer bound is a straightforward result of cut-set bound. In specific, \eqref{cut_set_user} is due to that the total number of data streams desired by each user cannot exceed the number of antennas $N$ it has, and \eqref{cut_set_BS} is due to that the total number of data streams sent from the BS cannot exceed the number of antennas $M$ it has.

\begin{theorem}[Inner Bound]\label{theorem2}
An inner bound of the DoF region of the $K$-user $(M,N)$ MIMO broadcast channel with general message sets, denoted as ${\mathcal D}_{\text{in}}$ is given by

 1) $\frac{M}{N}\in(0,1]$
  \begin{align}
   {\mathcal D}_{\text{in}}=\left\{{\bf d } \in {\mathbb R}^{2^K-1}_{+}: \sum\limits_{{\mathcal A}:{\mathcal A}\in \Psi} d_{\mathcal A}\leq M\right\};
  \end{align}

 2) $\frac{M}{N}\in(1,K]$
   \begin{align}
   {\mathcal D}_{\text{in}}=\text{conv}({\mathcal D}_1\cup{\mathcal D}_2),
  \end{align}
\quad with
  \begin{equation}\label{d_1}
    {\mathcal D}_1=\left\{{\bf d } \in {\mathbb R}^{2^K-1}_{+}: \sum\limits_{{\mathcal A}:i\in{\mathcal A}\in\Psi} d_{\mathcal{A}}\leq \frac{M}{K}, \quad \forall i\in[K]\right\},
  \end{equation}
 \begin{subequations}\label{achi_bound}
 \begin{align}
{\mathcal D}_2=\bigg\{&{\bf d} \in {\mathbb R}^{2^K-1}_{+}: \nonumber \\
                         &\sum\limits_{{\mathcal A}:i\in{\mathcal A}\in\Psi} d_{\mathcal{A}}\leq N,\quad \forall i\in[K]  \label{achi_bound_1}\\
                         &\sum\limits_{{\mathcal A}:{\mathcal A}\in \Psi} d_{\mathcal A}\leq M,  \label{achi_bound_2} \\
                         &{\bf 1}(d_{\mathcal{A}}>0)\bigg[(|\mathcal A|-1)\sum\limits_{{\mathcal B}:{\mathcal B}\in \Psi} d_{\mathcal B}+d_{\mathcal A}+\sum\limits_{{\mathcal B}:{\mathcal A}\subset{\mathcal B}\in\Psi}d_{\mathcal B}\bigg]\nonumber \\
                         &\quad\leq|\mathcal A|N,  \quad \ \forall {\mathcal A}\in\{{\mathcal A}:{\mathcal A}\in\Psi,|\mathcal A|\geq2\} \bigg\}; \label{achi_bound_3}
 \end{align}
 \end{subequations}

 3) $\frac{M}{N}\in(K,\infty)$
    \begin{align}
   {\mathcal D}_{\text{in}}=\left\{{\bf d } \in {\mathbb R}^{2^K-1}_{+}: \sum\limits_{{\mathcal A}:i\in{\mathcal A}\in\Psi} d_{\mathcal{A}}\leq N,\quad \forall i\in[K]\right\}.
  \end{align}
\end{theorem}
\begin{remark}[Optimality on two-user case]
  In the special case where $K=2$, by Theorem \ref{theorem2}, we have
  \begin{subequations} \label{DoF_two_user}
  \begin{align}
     {\mathcal D}_{\text{in}}=\Big\{&(d_{\{1\}},d_{\{2\}},d_{\{1,2\}}) \in {\mathbb R}^{3}_{+}: \nonumber \\
      &d_{\{1\}}+d_{\{1,2\}}\leq N,\label{DoF_two_user1} \\
      &d_{\{2\}}+d_{\{1,2\}}\leq N, \label{DoF_two_user2}\\
      &d_{\{1\}}+d_{\{2\}}+d_{\{1,2\}}\leq M, \label{DoF_two_user3}\\
      &{\bf 1}(d_{\{1,2\}}>0)(d_{\{1\}}+d_{\{2\}}+2d_{\{1,2\}})\leq 2N. \label{DoF_two_user4} \Big\}
  \end{align}
  \end{subequations}
  The inequalities \eqref{DoF_two_user1}, \eqref{DoF_two_user2} and \eqref{DoF_two_user3} are the cut-set bounds of DoF region. For the inequality \eqref{DoF_two_user4}, when $d_{\{1,2\}}=0$, it holds for any $d_{\{1\}}$, $d_{\{2\}}$ and $d_{\{1,2\}}$ due to the indicator function. When $d_{\{1,2\}}>0$, the inequality \eqref{DoF_two_user4} can be obtained by adding \eqref{DoF_two_user1} and \eqref{DoF_two_user2} and hence is redundant. As such, the inner bound in \eqref{DoF_two_user} coincides with  the outer bound and is thus optimal.
\end{remark}
\begin{remark}[Optimality on the general case]
  Comparing Theorem \ref{theorem1} with Theorem \ref{theorem2}, we see that the obtained DoF region is optimal when antenna configuration satisfies $\frac{M}{N}\in(0,1]\cup[K,\infty)$ at any number of users $K\geq2$. Moreover, the optimal DoF region for antenna configuration $\frac{M}{N}\in(0,1]$ only depends on $M$, indicating that further increasing the number of receive antennas at each user will not increase the DoF performance. Hence, $N-M$ receive antennas at each user are redundant and hence can be deactivated. Likewise, for the antenna configuration $\frac{M}{N}\in[K,\infty)$, we can activate only $KN$ transmit antennas at the BS without loss of optimality, yielding $M-KN$ redundant transmit antennas. The tightness of the inner and outer bounds for antenna configuration $\frac{M}{N}\in(1,K)$ shall be illustrated later in the NDT analysis.
\end{remark}

The rest of this section is dedicated to the proof of the inner bound using linear transmit precoding and receive combining schemes.

\subsection{Proof of Theorem \ref{theorem2}}
Define ${\bf s}_{\mathcal A}\in {\mathbb C}^{d_{\mathcal A}\times 1}$ as the signal vector intended for multicast group $\mathcal A$\footnote{In the case where $d_{\mathcal A}$ is not an integer, the achievable scheme needs $\kappa$-symbol extension such that $\kappa d_{\mathcal A}$ is an integer.}.  Define ${\bf U}_{\mathcal A}\in {\mathbb C}^{M\times d_{\mathcal A}}$ as the transmit precoding matrix of signal ${\bf s}_{\mathcal A}$. Define ${\bf V}^{i}_{\mathcal A}\in {\mathbb C}^{d_{\mathcal A} \times N}$ as the receive combining matrix of signal ${\bf s}_{\mathcal A}$ at user $i$, for $i\in \mathcal A$. The post-processed received signal at user $i$ can be expressed as
\begin{subequations}
  \begin{align}
  {\bf V}^{i}_{\mathcal A}{\bf y}_{i}&={\bf V}^{i}_{\mathcal A}{\bf H}_i{\bf x}+{\bf V}^{i}_{\mathcal A}{\bf n}_{i}\\
  &={\bf V}^{i}_{\mathcal A}{\bf H}_i{\bf U}{\bf s}+{\bf V}^{i}_{\mathcal A}{\bf n}_{i}\\
                   &={\bf V}^{i}_{\mathcal A}{\bf H}_i{\bf U}_{\mathcal A}{\bf s}_{\mathcal A}+{\bf V}^{i}_{\mathcal A}{\bf H}_i\sum\limits_{{\mathcal B} \in{\Psi}\setminus\{\mathcal A\} }{\bf U}_{\mathcal B}{\bf s}_{\mathcal B}+{\bf V}^{i}_{\mathcal A}{\bf n}_{i}
\end{align}
\end{subequations}
where
  \begin{subequations}
  \begin{align}
      &{\bf U}=\left[{\bf U}_{{\mathcal A}_1},{\bf U}_{{\mathcal A}_2},{\bf U}_{{\mathcal A}_3},\ldots,{\bf U}_{{\mathcal A}_{|\Psi|}}\right],\\
  &{\bf s}=\left[{\bf s}_{{\mathcal A}_1}^T,{\bf s}_{{\mathcal A}_2}^T,{\bf s}_{{\mathcal A}_3}^T,\ldots,{\bf s}_{{\mathcal A}_{|\Psi|}}^T\right]^T.
  \end{align}
  \end{subequations}

The design of all transmit precoding matrices and receive combining matrices depends on different antenna configurations, as shown below.

\subsubsection{$\frac{M}{N}\in\left (0,1 \right]$}
In this case, each user has enough signal dimensions to decode all the transmitted signals since the number of antennas at the user side exceeds that at the BS side. All the precoding matrices can be chosen randomly. Each user can use a zero-forcing based combining matrix to obtain their desired signals. Thus, the achievable DoF region is given by
\begin{align}
  {\mathcal D}_{\text{in}}=\left\{{\bf d } \in {\mathbb R}^{2^K-1}_{+}: \sum\limits_{{\mathcal A}:{\mathcal A}\in \Psi} d_{\mathcal A}\leq M\right\}.
\end{align}

\subsubsection{$\frac{M}{N}\in\left (K,\infty \right)$} \label{section_2}
 In this case, the number of antennas at the BS is larger than the total number of antennas at all the users. We concatenate all the signals to be decoded by each user $i$ as a meta-signal, i.e., $ {\bf{\hat s}}_i=\{{\bf s}_{\mathcal A}: i\in {\mathcal A} \in \Psi\}$, for all $i\in [K]$.  In this way, the broadcast channel with general message sets degenerates to the traditional broadcast channel with only unicast messages where ${\bf{\hat s}}_i$ is intended for user $i$ exclusively. A zero-forcing precoding matrix  can be used to cancel the inter-user interference.  All the combining matrices can be designed randomly. Thus, the achievable DoF region is given by
\begin{align}
   {\mathcal D}_{\text{in}}=\left\{{\bf d } \in {\mathbb R}^{2^K-1}_{+}: \sum\limits_{{\mathcal A}:i\in{\mathcal A}\in\Psi} d_{\mathcal{A}}\leq N,\quad \forall i\in[K]\right\}.
\end{align}

\subsubsection{$\frac{M}{N}\in\left (1,K \right]$}
The achievable DoF region in this case is the convex hull of ${\mathcal D}_1\cup{\mathcal D}_2$. The achievability of ${\mathcal D}_1$ in \eqref{d_1} can be proved by activating only $\frac{M}{K}$ receive antennas at each user\footnote{In the case where $\frac{M}{K}$ is not an integer, we use $\lfloor\frac{M}{K}\rfloor+1$ receive antennas at $M-K\lfloor\frac{M}{K}\rfloor$ users and use $\lfloor\frac{M}{K}\rfloor$ receive antennas at other users. By alternating user index and using time sharing, the DoF region ${\mathcal D}_1$ can be achieved.} and then applying the same achievable scheme as in the proof for antenna configuration $\frac{M}{N}\in\left (K,\infty \right)$ to achieve the region ${\mathcal D}_1$.

Then we give the achievable scheme of region ${\mathcal D}_2$ in \eqref{achi_bound}. The main idea is to design the receive combining matrices at user side to equalize the effective channels experienced by the same data stream on different target users and use a transmit zero-forcing precoding matrix to cancel inter-stream interference. In the following, we only activate $M_a\triangleq\sum\limits_{{\mathcal A}\in\Psi} d_{\mathcal A}$ transmit antennas at the BS. To ensure that any user $i$ in any multicast group ${\mathcal A}\in{\Psi}$ can decode the signal ${\bf s}_{\mathcal A}$, we can enforce the following sufficient conditions for the transmit precoding matrices $\{{\bf U}_{\mathcal A}\}$ and the receive combining matrices $\{{\bf V}^{i}_{\mathcal A}\}$:
\begin{subequations}
   \begin{align}
   &{\bf V}^{i}_{\mathcal A}{\bf H}_{i}{\bf U}_{\mathcal B}={\bf 0}, \hspace{0.8cm} \forall i\in{\mathcal A}, \  \forall {\mathcal A}\neq {\mathcal B}\in\Psi, \label{nulling}\\
   &\text{rank}({\bf V}^{i}_{\mathcal A}{\bf H}_{i}{\bf U}_{\mathcal A})=d_{\mathcal A}, \quad \forall i\in{\mathcal A}, \ \forall{\mathcal A}\in\Psi, \label{decode}
  \end{align}
\end{subequations}
where condition \eqref{nulling} is to force the interference caused by all other multicast messages to zero and condition \eqref{decode} is to ensure the decodability of the desired message ${\bf s}_{\mathcal A}$ with DoF $d_{\mathcal A}$. To satisfy the above conditions, we first equalize the effective channels experienced by the signal ${\bf s}_{\mathcal A}$ on all users in each multicast group ${\mathcal A}$ after multiplying their full-rank combining matrices, i.e.,
\begin{align} \label{effective_channel}
  {\bf V}^{i}_{\mathcal A}{\bf H}_i = {\bf G}_{\mathcal A}, \quad \forall i\in {\mathcal A}, \ \forall {\mathcal A}\in\Psi.
\end{align}
By doing so, the set of users in each multicast group can be seen as a single virtual user from the BS perspective. Then, the BS applies a zero-forcing based precoding matrix to send each multicast message, i.e.,
   \begin{align} \label{design_U}
     {\bf U}=\left[
               \begin{array}{c}
                 {\bf G}_{{\mathcal A}_1} \\
                 {\bf G}_{{\mathcal A}_2} \\
                 {\bf G}_{{\mathcal A}_3} \\
                      \vdots \\
                 {\bf G}_{{\mathcal A}_{|\Psi|}} \\
               \end{array}
             \right]^{\dag}.
   \end{align}

 In the following, we design the combining matrices $\{{\bf V}^i_{\mathcal A}\}$ to ensure that (i) all the combining matrices $\{{\bf V}^i_{\mathcal A}\}$ are full-rank, (ii) the condition \eqref{effective_channel} is satisfied for each multicast group ${\mathcal A}\in\Psi$ and (iii) the resulting ${\bf G}_{\mathcal A}$ is linearly independent to each other.

\begin{figure}[t]
    \begin{centering}
  \includegraphics[scale=0.6]{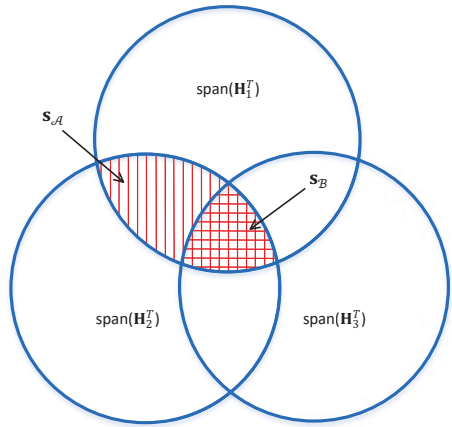}
   \caption{Illustration of the space of $\text{span}({\bf H}_i^T)$ where ${\mathcal A}=\{1,2\}$ and ${\mathcal B}=\{1,2,3\}$. It is clear that the space $\text{span}({\bf H}^{T}_1)\cap\text{span}({\bf H}^{T}_2)\cap\text{span}({\bf H}^{T}_3)$ is a subspace of $\text{span}({\bf H}^{T}_1)\cap\text{span}({\bf H}^{T}_2)$.} \label{space}
   \end{centering}
    \end{figure}

 For any multicast group $A\in \Psi$ that consists of $s\geq 2$ users, i.e.,
  \begin{equation}
    {\mathcal A}=\{i_1,i_2,\cdots,i_s\},
  \end{equation}
we can design the full-rank combining matrices as:
  \begin{align} \label{vs}
    &\left[
      \begin{array}{ccccc}
       {\bf G}_{{\mathcal A}} &  {\bf V}^{i_1}_{\mathcal A} & {\bf V}^{i_2}_{\mathcal A} &  \cdots & {\bf V}^{i_s}_{\mathcal A}\\
      \end{array}
    \right]^{T}\nonumber\\
    &\subseteq  \text{null}
    \left[
      \begin{array}{cccc}
        {\bf I}_{M_a} & {\bf I}_{M_a} & \cdots &  {\bf I}_{M_a}\\
        -{\bf H}_{i_1} & {\bf 0} & \cdots & {\bf 0} \\
        {\bf 0}   & -{\bf H}_{i_2} & \cdots& {\bf 0} \\
        \vdots & \vdots & \ddots & \vdots   \\
        {\bf 0} & {\bf 0} & {\bf 0} & -{\bf H}_{i_s}\\
      \end{array}
    \right]^{T},
  \end{align}
which meets the condition \eqref{effective_channel}. Now we analyze the linear independence between different ${\bf G}_{\mathcal A}$'s from the span space perspective.  From \eqref{vs}, we have:
   \begin{align}
     \text{span}({\bf G}^{T}_{\mathcal A})\subseteq\text{span}({\bf H}^{T}_{i_1})\cap\text{span}({\bf H}^{T}_{i_2})\cap\cdots\cap\text{span}({\bf H}^{T}_{i_s}).
   \end{align}
   This means that the row space of the effective channel matrix experienced by signal ${\bf s}_{\mathcal A}$ lies in the intersection space of $\{\text{span}({\bf H}^{T}_i)\}$ for all $i\in{\mathcal A}$. Thus, for the multicast groups ${\mathcal B}$'s where ${\mathcal B}\not\supset {\mathcal A}$ and $|\mathcal B|\geq2$, the effective channels $\{{\bf G}_{\mathcal B}\}$ are linearly independent to ${\bf G}_{\mathcal A}$.  Next, we turn to the case where ${\mathcal B}\supset {\mathcal A}$ and $|\mathcal B|\geq2$. Consider a multicast group ${\mathcal B}$ which includes all the users in ${\mathcal A}$ as well as an additional user $i_{s+1}$, i.e.,
   \begin{equation}
     {\mathcal B}=\{i_1,i_2,\cdots,i_s,i_{s+1}\}.
   \end{equation}
   Using the same method in \eqref{vs}, we can obtain the combining matrices $\{{\bf V}^i_{\mathcal B}\}$ and the resulting effective channel matrix ${\bf G}_{\mathcal B}$, which are given below for illustration convenience:
  \begin{align} \label{vs+1}
    &\left[
      \begin{array}{cccccc}
       {\bf G}_{\mathcal B} &  {\bf V}^{i_1}_{\mathcal B} & {\bf V}^{i_2}_{\mathcal B} &  \cdots & {\bf V}^{i_s}_{\mathcal B}& {\bf V}^{i_{s+1}}_{\mathcal B}\\
      \end{array}
    \right]^{T}\nonumber \\
    & \subseteq  \text{null}
    \left[
      \begin{array}{ccccc}
        {\bf I}_{M_a} & {\bf I}_{M_a} & \cdots &  {\bf I}_{M_a} &  {\bf I}_{M_a} \\
        -{\bf H}_{i_1} & {\bf 0} & \cdots & {\bf 0} & {\bf 0}  \\
        {\bf 0}   & -{\bf H}_{i_2} & \cdots& {\bf 0}  & {\bf 0} \\
        \vdots & \vdots & \ddots & \vdots & \vdots  \\
        {\bf 0} & {\bf 0} & {\bf 0} & -{\bf H}_{i_s} & {\bf 0}  \\
        {\bf 0} & {\bf 0} & {\bf 0} & {\bf 0} & -{\bf H}_{i_{s+1}}  \\
      \end{array}
    \right]^{T}.
  \end{align}
   Likewise, from \eqref{vs+1}, we have:
   \begin{align}
     &\text{span}({\bf G}^{T}_{\mathcal B})\nonumber \\
     &\subseteq\text{span}({\bf H}^{T}_{i_1})\cap\text{span}({\bf H}^{T}_{i_2})\cap\cdots\cap\text{span}({\bf H}^{T}_{i_s})\cap\text{span}({\bf H}^{T}_{i_{s+1}}).
   \end{align}
   This means that the row space of the effective channel matrix experienced by the multicast signal ${\bf s}_{\mathcal B}$ should lie in the intersection space of $\{\text{span}({\bf H}^{T}_i)\}$ for all $i\in{\mathcal B}$. Since ${\mathcal A}\subset{\mathcal B}$, the intersection space of $\{\text{span}({\bf H}^{T}_i)\}$ for all $i\in{\mathcal B}$ must be a subspace of the intersection space of $\{\text{span}({\bf H}^{T}_i)\}$ for all $i\in{\mathcal A}$. Fig. \ref{space} shows an example with ${\mathcal A}=\{1,2\}$ and ${\mathcal B}=\{1,2,3\}$. To assure the decodability, i.e., linear independence of ${\bf G}_{\mathcal A}$ and ${\bf G}_{\mathcal B}$, the total number of data streams which are intended for multicast groups ${\mathcal A}$ and ${\mathcal B}$ cannot exceed the dimension of the intersection space of $\{\text{span}({\bf H}^{T}_i)\}$ for all $i\in{\mathcal A}$, which is equal to the dimension of null space of the $sM_a\times (M_a+sN)$ matrix on right-hand side of \eqref{vs}. Using the null space theorem, we have:
   \begin{align}
     d_{\mathcal A}+d_{\mathcal B}\leq(M_a+sN)- sM_a.
   \end{align}
   Similarly, considering all the multicast groups $\mathcal B$'s where ${\mathcal B}\supset{\mathcal A}$, the total number of data streams which are intended for multicast groups ${\mathcal A}$ and all $\mathcal B$'s cannot exceed the dimension of the intersection space of $\{\text{span}({\bf H}^{T}_i)\}$ for all $i\in{\mathcal A}$. Using the null space theorem, we have:
   \begin{align} \label{indator_1}
     d_{\mathcal A}+\sum\limits_{{\mathcal B}\supset{\mathcal A}}d_{\mathcal B}\leq(M_a+sN)- sM_a.
   \end{align}
   Rearranging \eqref{indator_1} and noticing that \eqref{indator_1} is needed only when the signal ${\bf s}_{\mathcal A}$ is present, we have the equivalent condition shown in \eqref{achi_bound_3}. Thus, the linear independence of the effective channels $\{{\bf G}_{\mathcal B}\}$ where ${\mathcal B}\supset {\mathcal A}$ and $|\mathcal B|\geq2$ is guaranteed by \eqref{achi_bound_3}.

For each unicast message ${\bf s}_{\{i\}}$ with $i\in[K]$, they can be transmitted in the space of $\text{span}({\bf H}_i)$ and the corresponding combining matrix ${\bf V}^i_{\{i\}}$ can be designed randomly. By conditions \eqref{achi_bound_1} and \eqref{achi_bound_2}, the effective channels $\{{\bf G}_{\{i\}}\}$ of unicast messages are linearly independent of each other and other $\{{\bf G}_{\mathcal A}\}$, with ${\mathcal A}\in \Psi,|\mathcal A|\geq 2$.

As such, the region ${\mathcal D}_2$ is achievable. By using time sharing, the convex hull of ${\mathcal D}_1\cup{\mathcal D}_2$, i.e, ${\mathcal D}_{\text{in}}$ is achievable. Finally, Theorem \ref{theorem2} is proved.
 \begin{remark}[Complexity]
   The complexity of the above achievable scheme involves computing the receive combining matrices and the transmit precoding matrix and thus differs at different antenna configurations. In specific, when $\frac{M}{N}\in(0,1]$, the complexity is ${\emph O}(KM^3)$ since we need to compute $K$ zero-forcing based receive combining matrices, each with complexity ${\emph O}(M^3)$, and the complexity of designing the arbitrary transmit precoding matrix can be ignored. When $\frac{M}{N}\in(K,\infty)$, the complexity is ${\emph O}(K^3N^3)$ since the transmit precoding matrix is obtained by computing the inverse of a matrix ${\bf H}=[{\bf H}_1^T,{\bf H}_2^T,\cdots,{\bf H}_K^T]^T$ while the receive combining matrix of each user can be designed as an identity matrix with negligible complexity.  When $\frac{M}{N}\in(1,K]$,  we need to compute receive combining matrices for each of $2^K-1$ multicast groups based on \eqref{vs}, with complexity ${\emph O}(K^3N^3)$ for each multicast group, and compute the transmit precoding matrix based on \eqref{design_U} with complexity ${\emph O}(K^3N^3)$, yielding the total complexity of ${\emph O}(2^KK^3N^3)$.
 \end{remark}

\section{Achievable Minimum NDT}
Using the inner bound of the DoF region obtained in the previous section, we can obtain an achievable minimum NDT of the cache-aided MIMO broadcast channel for any given user demand at any given realization of cache placement. Recall that we shall deliver the set of coded messages $\{W_{\mathcal A}^{\oplus}\}$, each with length $a_{\mathcal A} F$ bits and intended for multicast group ${\mathcal A}\in \Psi$. Since the total delivery time is $T$, we have
\begin{align}
 R_{\mathcal A} T = a_{\mathcal A}F, \quad \forall {\mathcal A}\in{\Psi}. \label{rate}
\end{align}
where $R_A$ falls in the capacity region of the $K$-user $(M,N)$ MIMO broadcast channel with general message sets. By definition of NDT in \eqref{eqn NDTtau}, the above equation \eqref{rate} can be rewritten asymptotically when $F\rightarrow \infty$ and $P\rightarrow \infty$ as
\begin{align}\label{time}
  d_{\mathcal A} \tau=a_{\mathcal A}, \quad \forall {\mathcal A}\in{\Psi}.
\end{align}

Thus, finding the achievable minimum NDT for a given set of coded messages $\{W_{\mathcal A}^{\oplus}\}$ with message length vector $\bf f$ is equivalent to finding a DoF tuple $\bf d$ within the inner bound satisfying \eqref{time} at the minimum possible $\tau$. This can be formulated as an optimization problem below:
  \begin{subequations}
    \begin{align}
  \mathcal{P}_1: \quad \tau_a(\bf f) \triangleq &\min\limits_{\tau,\bf d} \ \tau  \\
  &\ \text{s.t. } \ \tau{\bf d}={\bf f}, \quad  \label{P1_time}\\
  &\quad \quad  \ \; {\bf d}\in {\mathcal D}_{\text{in}},\label{P1_region}
 \end{align}
  \end{subequations}
where ${\mathcal D}_{\text{in}}$ is given in Theorem \ref{theorem2}.

The optimal solution of ${\mathcal P}_1$ satisfies that the DoF tuple ${\bf d}={\bf f}/\tau_a$ lies on the boundary of ${\mathcal D}_{\text{in}}$. Assume that there are $Z$ corner points in ${\mathcal D}_{\text{in}}$, each denoted as ${\bf e}_j$, for $j\in[Z]$. Since any boundary point of ${\mathcal D}_{\text{in}}$ can be written as a convex combination of all the corner points, the problem ${\mathcal P}_1$ can be equivalently represented as
  \begin{subequations}
    \begin{align}
     \mathcal{P}_2: \quad \tau_a(\bf f)= &\min \ \sum\limits_{j=1}^{Z} \beta_j  \\
     &\ \text{s.t. } \ \sum\limits_{j=1}^{Z} \beta_j{\bf e}_j={\bf f}, \quad \\
     &\quad \ \quad  \beta_j\geq0, \quad \forall j\in[Z].
    \end{align}
  \end{subequations}
Problem ${\mathcal P}_2$ is a linear programming problem and can be solved efficiently by some linear equation substitution and other manipulations.

\begin{remark}[Finding corner points]\label{tdd_optimal}
 The corner points of the DoF region can be found by using the algorithm in Appendix \ref{corner_point}. In the special case with  $\frac{M}{N}\in(0,1]$, the corner points are the DoF tuples each of which has one element being $M$ and others being zero. This means that the optimal solution of ${\mathcal P}_1$ when $\frac{M}{N}\in(0,1]$ is to deliver different coded messages sequentially in a time-sharing manner.
 \end{remark}

Next, we give an example at a specific message length vector $\bf f$ to illustrate the solution of problem ${\mathcal P}_1$ as well as effectiveness of the proposed content delivery scheme.

\textbf{Example 1}. Consider a three-user $(M = 5,N = 3)$ cache-aided MIMO broadcast channel and a library of $L=4$ files. The normalized cache size is $\mu=0.4$. The lengths of cached subfiles for a specific realization of decentralized cache placement are given in Table \ref{table1}. In the delivery phase, we assume the worst-case user demand ${\bf r}=(1,2,3)^T$. By using the coding method in \cite{long_glo17}, the length vector of coded messages is given by ${\bf f}=[a_{\{1\}}=\frac{1}{5},a_{\{2\}}=\frac{1}{10},a_{\{3\}}=0,a_{\{1,2\}}=\frac{3}{20},a_{\{1,3\}}=\frac{1}{4},a_{\{2,3\}}=\frac{7}{20},a_{\{1,2,3\}}=0]$. Using our proposed delivery scheme, by solving ${\mathcal P}_1$, we obtain $\tau_a=\frac{7}{30}$ and the corresponding DoF tuple is ${\bf d}^*=[d_{\{1\}}=\frac{6}{7},d_{\{2\}}=\frac{3}{7},d_{\{3\}}=0,d_{\{1,2\}}=\frac{9}{14},d_{\{1,3\}}=\frac{15}{14},d_{\{2,3\}}=\frac{3}{2},d_{\{1,2,3\}}=0]$. For comparison, we also consider a conventional delivery scheme where these coded messages are transmitted one by one. Since the DoF of the channel when sending each message separately is given by $3$, we have the achievable NDT as $\tau=\frac{\frac{1}{5}+\frac{1}{10}+\frac{3}{20}+\frac{1}{4}+\frac{7}{20}}{3}=\frac{7}{20}$. It can be seen that our achievable NDT is only 66.7\% of that of the benchmark scheme. This example clearly shows that higher spatial multiplexing gain of MIMO is exploited to transmit multiple coded messages simultaneously.
 \begin{table*}
 \centering
 \caption{The specific decentralized cache placement in Example 1.}\label{table1}
    \begin{tabular}{|c|c|c|c|c|c|c|c|c|}
   \hline
   \diagbox{$\ell$}{($F$ bits)}{$|W_{\ell,\mathcal A}|$} & $|W_{\ell,\{1\}}|$ & $|W_{\ell,\{2\}}|$ & $|W_{\ell,\{3\}}|$ & $|W_{\ell,\{1,2\}}|$ & $|W_{\ell,\{1,3\}}|$ & $|W_{\ell,\{2,3\}}|$ & $|W_{\ell,\{1,2,3\}}|$ & $|W_{\ell,\varnothing}|$ \\
   \hline
   1 & 0.05 & 0.1  & 0.25 & 0.25 & 0.1  & 0.05 & 0    & 0.2 \\
   \hline
   2 & 0.15 & 0.3  & 0.2  & 0.05 & 0.15 & 0    & 0.05 & 0.1 \\
   \hline
   3 & 0.25 & 0.35 & 0.2  & 0    & 0.15 & 0.05 & 0    & 0   \\
   \hline
   4 & 0.2  & 0.15 & 0.15 & 0.05 & 0.05 & 0.1  & 0.1  & 0.2 \\
   \hline
 \end{tabular}
 \end{table*}

 Note that ${\mathcal P}_1$ is defined for any given set of coded messages with length vector $\bf f$ and hence can be computed for any user demand at any realization of cache placement, no matter it is centralized or decentralized. In some special cases where the lengths of coded messages can be expressed explicitly, we can obtain closed-form expressions for the achievable minimum NDT of the system. This leads to the following useful Corollaries whose proofs can be found in Appendices \ref{proof_of_corollary_1} and \ref{proof_of_corollary_2}, respectively.

\begin{corollary}\label{ndt_centralized}
 If the centralized cache placement as in \cite{Alilimits} is adopted, the achievable minimum NDT for the worst-case user demand at normalized caching size $\mu\in\left\{0,\frac{1}{K},\frac{2}{K},\ldots,1\right\}$ is given by
\begin{align}\label{our_tau}
&\tau_{\text{worst}}\nonumber \\
           &=\begin{cases}
           \frac{K(1-\mu)}{M(K\mu+1)},  \hspace{22mm} \frac{M}{N}\in\left(0,\frac{(K\mu+1)\binom{K}{K\mu+1}}{1+K\mu\binom{K}{K\mu+1}}\right]\\
           \min\left\{\frac{K(1-\mu)}{M}, \frac{1-\mu}{N(K\mu+1)}\left[\frac{1}{\binom{K-1}{K\mu}}+\frac{K^2\mu}{(K\mu+1)}\right]\right\},\\
            \hspace{37mm}\frac{M}{N}\in\left(\frac{(K\mu+1)\binom{K}{K\mu+1}}{1+K\mu\binom{K}{K\mu+1}},K\right] \\
           \frac{1-\mu}{N}, \hspace{29mm}\frac{M}{N}\in(K,\infty) \\
           \end{cases}.
\end{align}
For the general case where $K\mu$ is not an integer, the lower convex envelope of the above points is achievable.
\end{corollary}

\begin{remark}[Discussion on centralized caching]\label{comparasion}
 In this remark, we compare Corollary \ref{ndt_centralized} with existing results on centralized caching. First, for the special case with $K=3$ users, the above minimum NDT is the same as the result in \cite{long_journal} where the three-user MIMO broadcast channel is obtained by letting the normalized cache size of each transmitter in the $3\times 3$ cache-aided MIMO interference network be one. Next, for the special case with single antenna at each user, i.e., $N=1$, the following achievable NDT for the worst-case user demand can be obtained from \cite{Caire2017multi} and \cite{MISO_Petros_Elia} as\footnote{In \cite{MISO_Petros_Elia}, \eqref{tau_benchamrk} only holds when $M$ is a divisor of $K$.}
  \begin{align}\label{tau_benchamrk}
    \tau'_{\text{worst}}=\frac{1-\mu}{\min\left\{1,\frac{K\mu+M}{K}\right\}},
  \end{align}
  for $\mu\in\left\{0,\frac{1}{K},\frac{2}{K},\ldots,1\right\}$. By comparing \eqref{our_tau} and \eqref{tau_benchamrk}, it can be seen that $\tau_{\text{worst}}=\tau'_{\text{worst}}$ when (i) $K=3$, or (ii) $M\in\left[1,\left\lfloor\frac{K(K\mu+1)\binom{K-1}{K\mu}}{1+K\mu\binom{K}{K\mu+1}}\right\rfloor-K\mu\right]$ with $\mu\leq 1-\frac{M}{K}$, or (iii) $M\in[K,\infty)$. For other cases, we have $\tau_{\text{worst}}>\tau'_{\text{worst}}$. This performance loss is due to that our scheme is motivated by decentralized caching where each of the coded messages to be delivered can be of arbitrary length and for arbitrary user set. Inevitably, it may suffer when these messages have the same length and are intended to symmetric user sets as it is for centralized caching and worst-case user demand.
\end{remark}

\begin{corollary}
   If the decentralized cache placement as in \cite{Alidecentralized} is adopted and the file length $F$ is large enough, the achievable minimum NDT for the worst-case user demand at any normalized cache size $\mu$ is given by
 \begin{equation}\label{NDT_de}
  \tau_{\text{worst}}=
            \begin{cases}
                \sum\limits_{s=1}^{K}\frac{\mu^{s-1}(1-\mu)^{K-s+1}\binom{K}{s}}{M} , & \frac{M}{N}\in(0,1]\\
                \tau_1   , & \frac{M}{N}\in(1,K] \\
                \frac{1-\mu}{N},& \frac{M}{N}\in(K,\infty) \\
           \end{cases}.
\end{equation}
where $\tau_1$ is the optimal solution of problem ${\mathcal P}_1$ with each element of $\bf f$ being $a_{\mathcal A}=\mu^{|\mathcal A|-1}(1-\mu)^{K-|\mathcal A|+1}$ and ${\mathcal D}_{\text{in}}=\text{conv}({\mathcal D}_1\cup\tilde{{\mathcal D}}_2)$. Here, $\tilde{{\mathcal D}}_2$ is given by:
\begin{subequations}\label{simplified_d}
\begin{align}
\tilde{{\mathcal D}}_2=\Big\{&{\bf d } \in {\mathbb R}^{K}_{+}| \nonumber \\
&\sum\limits_{s=1}^{K} \binom{K-1}{s-1}d_s\leq N, \\
&\sum\limits_{s=1}^{K} \binom{K}{s}d_s\leq M, \\
&{\bf 1}(d_s> 0)\bigg[(s-1)\sum\limits_{s'=1}^{K}\binom{K}{s'}d_{s'}+d_s\nonumber \\
&\ \,+\sum\limits_{s'=s+1}^{K}\binom{K-s}{s'-s}d_{s'}\bigg]\leq sN, \ \forall s\in\{2,\cdots,K\}\Big\}.
\end{align}
\end{subequations}
\end{corollary}
\begin{remark}[Large number of transmit antennas]\label{remark6}
   It is observed from Corollary 1 and 2 that the worst-case NDT of centralized and decentralized caching is the same and given by $\frac{1-\mu}{N}$ when antenna configuration satisfies $\frac{M}{N} \in [K, \infty)$. It is easy to prove that $\frac{1-\mu}{N}$ is also the converse of the minimum NDT for any caching and delivery strategy, since $1-\mu$ is the uncached fraction of the desired file and $N$ is the maximum number of data streams each user can receive. Further, it is seen from the proof of Theorem \ref{theorem2} when $\frac{M}{N} \in [K, \infty)$ that the broadcast channel with general message sets degenerates to the broadcast channel with private (unicast) messages only. As such, we can conclude that when the number of transmit antennas is large enough, i.e., $M\geq KN$, (i) both centralized and decentralized cache placements are globally optimal, and (ii) the spatial multiplexing gain overwrites the coded multicasting gain.
\end{remark}

\section{Optimality Analysis}\label{section_optimality_analysis}
 This section is to show the (order-)optimality of the proposed MIMO transmission scheme of treating the content delivery in multi-antenna coded caching as general message sets. We first present a lower bound of the minimum NDT by using the outer bound of DoF region and then analyse the multiplicative gap between the achievable minimum NDT and the lower bound.

A lower bound of the minimum NDT for a given set of coded messages with message length vector $\bf f$, can be given by the optimal solution of the following problem:
 \begin{subequations}
 \begin{align}
  \mathcal{P}_3: \quad \tau_l(\bf f)  \triangleq &\min\limits_{\bf d} \ \tau  \\
  &\ \text{s.t. } \ \tau{\bf d}={\bf f}, \quad  \\
  &\quad \quad  \ \; {\bf d}\in {\mathcal D}_{\text{out}}. \label{out_cons}
  \end{align}
 \end{subequations}
 Similar to solving ${\mathcal P}_1$, finding the optimal solution of $\mathcal P_3$ is to find a $\tau_l$ such that ${\bf f}/\tau_l$ is on the boundary of ${\mathcal D}_{\text{out}}$.  Note that this lower bound of NDT for a given set of coded messages should not be confused with the converse of the minimum NDT for all achievable caching and delivery schemes as studied in \cite{long_journal,xu,simeone}. The lower bound in this paper is mainly to show the impact of the tightness of the achievable DoF region on the delivery time for a given set of coded messages. The multiplicative gap between our achievable minimum NDT and the lower bound is given in Theorem \ref{theorem_gap}.
\begin{theorem}\label{theorem_gap}
  The achievable minimum NDT for a given set of coded messages $\{W_{\mathcal A}^{\oplus}\}$ is optimal when the antenna configuration satisfies $\frac{M}{N}\in(0,1]\cup[K,\infty)$ and is within a multiplicative factor of $\frac{M}{N}$ to the optimum when $\frac{M}{N}\in(1,K)$.
\end{theorem}
\begin{IEEEproof}
 We first prove the optimality for antenna configuration $\frac{M}{N}\in(0,1]\cup[K,\infty)$. The inner bound ${\mathcal D}_{\text{in}}$ coincides with the outer bound ${\mathcal D}_{\text{out}}$ in these regions. Thus, the solution of the problem $\mathcal{P}_3$ is the same as the one of the problem $\mathcal{P}_1$.

  Next, we prove the multiplicative gap between the achievable NDT and the lower bound when $\frac{M}{N}\in(1,K)$. We remove the indicator functions of bounds \eqref{achi_bound_3} in region ${\mathcal D}_2$ and denote the subset of ${\mathcal D}_2$ as $\bar{{\mathcal D}}_2$. Since the DoF region ${\mathcal D}_{\text{in}}$ for the antenna configuration $\frac{M}{N}\in(1,K)$ is $\text{conv}({\mathcal D}_1,{\mathcal D}_2)$, $\bar{{\mathcal D}}_2$ is also a subset of ${\mathcal D}_{\text{in}}$. Hence, the optimal solution of the following problem can be regarded as an upper bound of our achievable minimum NDT:
 \begin{subequations}
    \begin{align}
  \mathcal{P}_4: \quad \tau_u(\bf f)&\  \triangleq \ \min\limits_{\bf d} \ \tau  \\
  &\ \text{s.t. } \ \tau{\bf d}={\bf f}, \quad  \\
  &\quad \quad  \ \; {\bf d}\in \bar{{\mathcal D}}_2.  \label{up_cons}
 \end{align}
 \end{subequations}
For convenience, we rewrite the constraint \eqref{up_cons} as the following form
\begin{align}
{\bf A}_{1}{\bf d}\leq {\bf c}_{1},
\end{align}
where the elements of ${\bf A}_{1}$ and ${\bf c}_{1}$ are constants for any given antenna configuration $(M,N)$. The optimal solution of problem $\mathcal{P}_4$ can be represented as
\begin{align}
  \tau_{u}\triangleq\parallel{\bf A}_{1}{\bf f}./{\bf c}_{1}\parallel_{\infty}.
\end{align}
Similarly, we rewrite the constraint \eqref{out_cons} as the following form
\begin{align}
{\bf A}_{2}{\bf d}\leq {\bf c}_{2},
\end{align}
where the elements of ${\bf A}_{2}$ and ${\bf c}_{2}$ are constants for any given antenna configuration $(M,N)$. The lower bound of NDT, i.e., the optimal solution of problem $\mathcal{P}_3$ can be represented as
\begin{align}
  \tau_{l}\triangleq\parallel{\bf A}_{2}{\bf f}./{\bf c}_{2}\parallel_{\infty}.
\end{align}
Thus, the multiplicative gap between our achievable NDT and the lower bound is less than $\frac{M}{N}$ as shown in \eqref{gap}.
\begin{align}\label{gap}
  \rho&\triangleq\frac{\tau_a}{\tau_l} \nonumber \\
     &\leq\frac{\tau_u}{\tau_l}\nonumber \\
      &=\frac{\parallel{\bf A}_{1}{\bf f}./{\bf c}_{1}\parallel_{\infty}}{\parallel{\bf A}_{2}{\bf f}./{\bf c}_{2}\parallel_{\infty}}\nonumber \\
   &=\frac{1}{\parallel{\bf A}_{2}{\bf f}./{\bf c}_{2}\parallel_{\infty}}\max\bigg\{\parallel{\bf A}_{2}{\bf f}./{\bf c}_{2}\parallel_{\infty},\nonumber \\
   &\hspace{13mm}\max\limits_{\mathcal A}\frac{1}{|\mathcal A|N}\bigg[(|\mathcal A|-1)M_a+d_{\mathcal A}+\sum\limits_{{\mathcal B}:{\mathcal A}\subset{\mathcal B} \in\Psi }d_{\mathcal B}\bigg]\bigg\} \nonumber \\
  &\leq \frac{1}{\parallel{\bf A}_{2}{\bf f}./{\bf c}_{2}\parallel_{\infty}}\max\left\{\parallel{\bf A}_{2}{\bf f}./{\bf c}_{2}\parallel_{\infty},M_a/N\right\}\nonumber \\
    &=\max\left\{1,\frac{M_a/N}{\max\left\{M_a/M,\frac{1}{N}\max\limits_{i}\sum\limits_{{\mathcal A}:i\in{\mathcal A}\in\Psi}d_{\mathcal A}\right\}}\right\} \nonumber \\
    &\leq\frac{M}{N}.
\end{align}
\end{IEEEproof}

  Although the achievable DoF region derived in this paper for the MIMO broadcast channel with general message sets is not globally optimal, Theorem \ref{theorem_gap} implies that the globally optimal delivery scheme cannot provide any gain more than a bounded gap in terms of the NDT.

\section{Numerical Results and Discussions}
 In this section, we present some numerical results of our proposed content delivery scheme for multi-antenna coded caching. Throughout this section, we consider a library with $L = 4$ files, each of length $F = 100$ bits. We show the average NDT results among different user demands for both centralized caching and decentralized caching. In specific, for centralized caching\footnote{We calculate the average NDT at the normalized cache size $\mu\in\left\{0,\frac{1}{K},\frac{2}{K},\ldots,1\right\}$. For a general case where $K\mu$ is not an integer, the average NDT is the lower convex envelope of these points by using memory sharing \cite{Alilimits}.}, we consider all the possible $4^K$ user demands and use the algorithm in \cite{yu2016exact}, which can exploit commonality of user demands and is optimal under uncoded prefetching in one-server shared link, to generate coded messages. For decentralized caching, we consider 1000 independent realizations of cache placement, where for each cache realization each user requests any of the four files with equal probability, and the algorithm in \cite{naifu} is used to generate coded messages.

  To illustrate the advantage of the proposed content delivery method based on general message sets, we consider the following two benchmark schemes for comparison:
  \begin{itemize}
  \item Time sharing: Each generated coded message is sent one by one with per-message DoF being $\min\{M,N\}$.
  \item Group-by-group: All the coded messages are divided into groups according to the sizes of intended user sets. That is, those coded messages intended to the same number of users are grouped together. In case they have different message lengths, zero-padding is applied. The coded messages are then delivered group by group. Within each group, the transmission method in \cite{Caire2017multi} is adopted. Note that the method in \cite{Caire2017multi} is limited to single receive antenna only with $N=1$.
  \end{itemize}

\begin{figure}[t]
    \begin{centering}
  \includegraphics[scale=0.46]{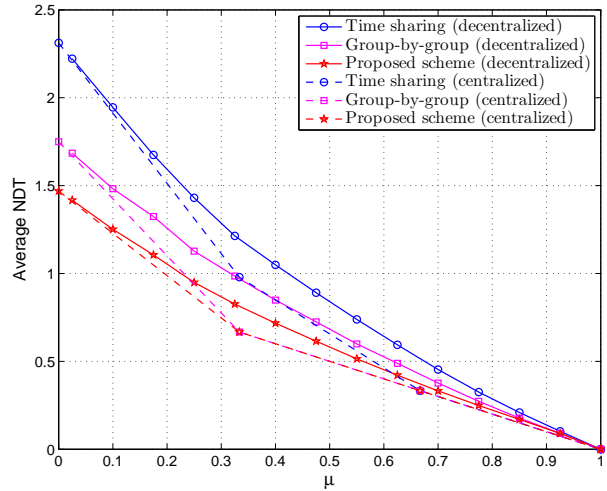}
   \caption{The average NDT of the proposed scheme and benchmark schemes with $K=3,M=2,N=1$.} \label{simulation1}
   \end{centering}
    \end{figure}

Fig. \ref{simulation1} shows the average NDT results versus normalized cache size in the system with $K=3$, $M=2$ and $N=1$. It can be generally observed that the proposed scheme outperforms the two benchmark schemes. In specific, for decentralized caching, the proposed scheme has superior performance in the whole cache size region. This is because our proposed delivery scheme can transmit coded messages intended for unequal-size multicast groups concurrently to fully exploit spatial multiplexing gain. For centralized caching, the proposed scheme performs the best when $\mu<\frac{1}{3}$. Again, the performance improvement comes from the simultaneous transmission of coded messages intended for unequal-size multicast groups. Note that the unequal-size multicast groups in centralized caching are due to that we adopted the coding method in \cite{yu2016exact} which takes into account the commonality of user demands. When $\frac{1}{3}\leq\mu<\frac{2}{3}$, the proposed scheme and group-by-group scheme perform the same since the intended multicast groups are symmetric with the same size. When $\mu\geq\frac{1}{3}$, the three schemes achieve the same performance because for any user demand, there is only one coded message to be delivered and it is intended for all the users.

 \begin{figure}[t]
    \begin{centering}
  \includegraphics[scale=0.46]{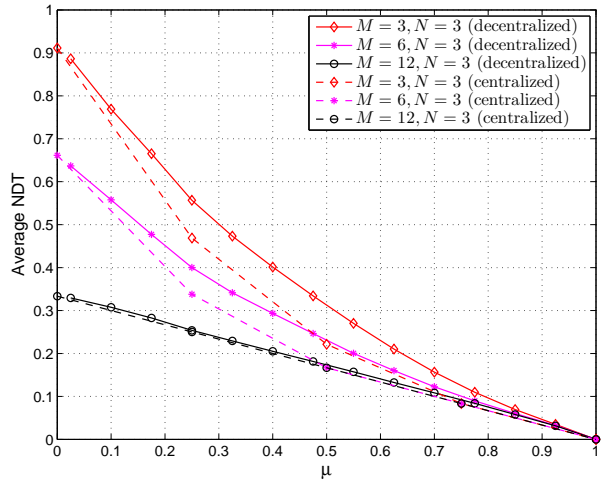}
   \caption{The average NDT of the proposed scheme at different antenna configurations with $K=4$.} \label{simulation2}
   \end{centering}
    \end{figure}

Next, we demonstrate the effect of different antenna configurations. Fig. \ref{simulation2} plots the average NDT results when the normalized cache size varies in a system with $K=4$ users. We can see that the average NDT of the proposed scheme decreases with the number of transmit antennas $M$. It is observed that the centralized caching outperforms the decentralized caching when $M=3, 6$ and they are nearly the same when $M=12$. The close performance between them at $M=12$ is due to that the spatial multiplexing gain is large enough to overwrite the coded multicasting gain. Similar observation also holds for the worst-case NDT performance as discussed in Remark \ref{remark6}.

Finally, we illustrate the actual delivery time of the proposed scheme in the finite SNR region. The channel is assumed to be normalized Rayleigh fading. We use the setting in Example 1 in Section IV for simulation, where the cache placement is given in Table \ref{table1} and the worst-case user demand is considered. Note that the DoF tuple ${\bf d}^*$ for this given example is achieved by using time sharing of three corner points in the DoF region ${\mathcal D}_{\text{in}}$, i.e, ${\bf d}^*=\frac{2}{7}{\bf d}_1+\frac{2}{7}{\bf d}_2+\frac{3}{7}{\bf d}_3$, where ${\bf d}_1=[3,0,0,0,0,\frac{3}{2},0]$, ${\bf d}_2=[0,\frac{3}{2},0,0,\frac{3}{2},\frac{3}{2},0]$, and ${\bf d}_3=[0,0,0,\frac{3}{2},\frac{3}{2},\frac{3}{2},0]$. This means that the total transmission takes three phases. In phase 1, we deliver coded messages $\{W^{\oplus}_{\{1\}},W^{\oplus}_{\{2,3\}}\}$ with length vector
${\bf f}_1=\frac{2}{7}\times\tau_a\times{\bf d}_1=\left[\frac{1}{5},0,0,0,0,\frac{1}{10},0\right]$.
The corresponding delivery time $T_1$ is computed as
\begin{align}
 T_1=\max\left\{\frac{F/5}{\bar{R}_{\{1\}}(P)},\frac{F/10}{\bar{R}_{\{2,3\}}(P)}\right\},
\end{align}
where $\bar{R}_{\mathcal A}(P)$ is the average transmission rate for message $W^{\oplus}_{\mathcal A}$ with transmit power $P$. In specific, $\bar{R}_{\{1\}}(P)$ is given by
\begin{align}
  &\bar{R}_{\{1\}}(P)\nonumber \\
  &=\mathbb{E}_{\bf H}\log\left( \det\left[{\bf I}+P{\bf V}^1_{\{1\}}{\bf H}_1{\bf U}_{\{1\}}({\bf V}^1_{\{1\}}{\bf H}_1{\bf U}_{\{1\}})^H\right] \right),
\end{align}
where the normalized combining matrix ${\bf V}^1_{\{1\}}$ is generated by the standard normal distribution and the normalized precoding matrix ${\bf U}_{\{1\}}$ is designed by \eqref{design_U}, and $\bar{R}_{\{2,3\}}(P)$ is given by
\begin{align}
  &\bar{R}_{\{2,3\}}(P)=\mathbb{E}_{\bf H}\min\{{R}^2_{\{2,3\}}(P),{R}^3_{\{2,3\}}(P)\},
\end{align}
with
\begin{align}
  &{R}^2_{\{2,3\}}(P)\nonumber \\
  &=\log\left( \det\left[{\bf I}+P{\bf V}^2_{\{2,3\}}{\bf H}_2{\bf U}_{\{2,3\}}({\bf V}^2_{\{2,3\}}{\bf H}_2{\bf U}_{\{2,3\}})^H\right] \right),\\
  &{R}^3_{\{2,3\}}(P)\nonumber \\
  &=\log\left( \det\left[{\bf I}+P{\bf V}^3_{\{2,3\}}{\bf H}_3{\bf U}_{\{2,3\}}({\bf V}^3_{\{2,3\}}{\bf H}_3{\bf U}_{\{2,3\}})^H\right] \right),
\end{align}
where ${\bf V}^2_{\{2,3\}}$ and ${\bf V}^3_{\{2,3\}}$ are designed by \eqref{vs} and ${\bf U}_{\{2,3\}}$ is designed by \eqref{design_U}. The delivery time of phases 2 and 3 can be computed similarly. Summing up the delivery time of the three phases, we can obtain the total delivery time $T$.

\begin{figure}[t]
    \begin{centering}
  \includegraphics[scale=0.46]{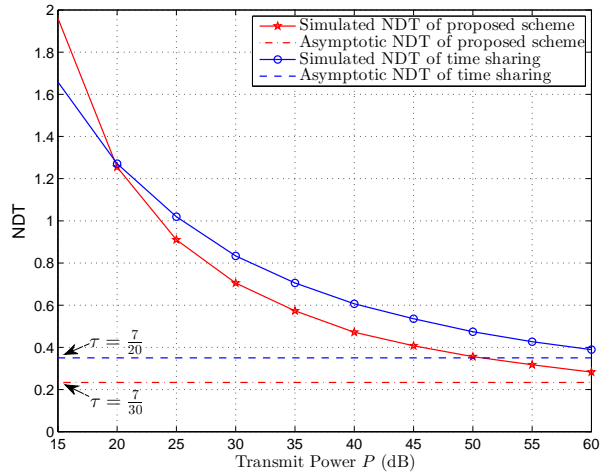}
\caption{Comparison between simulated NDT and asymptotic NDT of the proposed scheme under the setting in Example 1.}\label{simulation_finite}
   \end{centering}
    \end{figure}

Fig. \ref{simulation_finite} shows the simulated NDT, computed as $\frac{T}{F/\log P}$, versus transmit power $P$ (normalized by unit noise variance) for Example 1. Here, we only choose the time sharing scheme as benchmark since the group-by-group scheme is not applicable. Each result is averaged over 1000 independent channel realizations. It can be seen that the simulated NDT approaches the asymptotic NDT when the transmit power is large enough for both schemes, which verifies the theoretical analysis in this paper. It is also observed that the time sharing scheme has better performance at low SNR region ($P<20 \text{dB}$) while our proposed scheme is superior at high SNR region ($P\geq 20\text{dB}$). This is because our proposed scheme is designed to optimize the asymptotic performance rather than the actual delivery time. Further investigation is needed to improve the performance at low SNR but beyond the scope of this paper.

\section{Conclusion}
In this paper, we proposed to treat the content delivery in multi-antenna coded caching as transmission with general message sets. We derived an achievable DoF region of MIMO broadcast channel with general message sets. The specific design of transmit precoding and receive combining is based on different antenna configurations. By using the achievable DoF region, an upper bound of the minimum NDT is obtained for any given user demand at any given realization of cache placement. The achievable NDT illustrates the spatial multiplexing gain of MIMO through simultaneous transmission of multiple coded messages intended to different multicast groups. We also presented a lower bound of the NDT by using the cut-set outer bound of the DoF region. It is shown that the achievable NDT is optimal at certain cases and is within a bounded gap from the optimum at other cases.

\begin{appendices}
\section{The Algorithm of Finding Corner Points}\label{corner_point}
\subsection{$\frac{M}{N}\in(0,1]$}
For the antenna configuration $\frac{M}{N}\in(0,1]$, each of the corner points takes the form
\begin{equation}
  {\bf e}_{j}=\{0,0,\cdots,\underbrace{M}_{\text{$j$-th element}},\cdots,0\},   \quad \forall j\in[2^K-1].
\end{equation}

\subsection{$\frac{M}{N}\in(K,\infty)$}
For the antenna configuration $\frac{M}{N}\in(K,\infty)$, considering a set of multicast groups $\{\mathcal A_j\}_{j=1}^{C}$, the DoF tuple is a corner point if the following conditions are satisfied:
\begin{enumerate}
  \item $d_{\mathcal A_j}=N$, $\forall j\in[C]$ and other elements are zero,
  \item ${\mathcal A}_j\cap{\mathcal A}_{j'}=\varnothing, \forall j\neq j'\in[C]$.
\end{enumerate}

\subsection{$\frac{M}{N}\in(1,K]$}
Since the DoF region ${\mathcal D}_{\text{in}}$ for the antenna configuration $\frac{M}{N}\in(1,K]$ is $\text{conv}({\mathcal D}_1\cup{\mathcal D}_2)$, the corner points of ${\mathcal D}_1$ and ${\mathcal D}_2$ contain all the corner points of ${\mathcal D}_{\text{in}}$. We first consider the region ${\mathcal D}_1$. The corner points of ${\mathcal D}_1$ take the similar form as the ones for antenna configuration $\frac{M}{N}\in(K,\infty)$. Then we consider the region ${\mathcal D}_2$. For convenience, we define an one to one mapping to simplify the indices of multicast groups:
\begin{align}
  &\{1\}\leftrightarrow 1, \nonumber \\
  &\{2\}\leftrightarrow 2, \nonumber \\
     & \quad   \vdots          \nonumber \\
  &[K]\leftrightarrow 2^K-1. \nonumber
\end{align}
The multicast group with less cardinality is mapped with a smaller number. If the multicast groups have the same cardinality, the multicast group which has smaller user index is mapped with a smaller number. We denote the bounds \eqref{achi_bound_1} and \eqref{achi_bound_2} in ${\mathcal D}_2$ as $\text{C}_1,\cdots,\text{C}_K$, and $\text{C}_{2^K}$ respectively. We remove the indicator functions in bounds \eqref{achi_bound_3} and denote these processed bounds as $\text{C}_{K+1},\cdots,\text{C}_{2^K-1}$, respectively. The method to find the corner points of ${\mathcal D}_2$ is given in Algorithm 1.

\begin{algorithm}\label{algo_1}
\caption{Finding corner points}
\begin{algorithmic}[1]
\FORALL{$i=1:2^K-1$}
     \FORALL{$\Gamma\subseteq[2^K-1]$ with $|\Gamma|=i$}
         \STATE set $d_j=0$ in the DoF tuple, $\forall j\in[2^K-1]\setminus\Gamma$.
        \FORALL{$\Lambda\subseteq[2^{K}]\setminus\Gamma\cup[K]$ with $|\Lambda|=i$ }
             \STATE calculate the intersection point of the bounds $\{\text{C}_p\}$, $\forall p\in\Lambda$ in ${\mathbb R}^{i}_{+}$ and denote the intersection point as ${\bf d}_{\Gamma,\Lambda}$.
               \IF{${\bf d}_{\Gamma,\Lambda}$ is within the bounds $\{\text{C}_p\}$, $\forall p\in[2^K]\setminus(\Gamma\cup\Lambda)$}
                   \STATE ${\bf d}_{\Gamma,\Lambda}$ is a corner point.
               \ENDIF
       \ENDFOR
     \ENDFOR
\ENDFOR
\end{algorithmic}
\end{algorithm}

\section{Proof of Corollary 1}\label{proof_of_corollary_1}
In this section, we present the proof of Corollary 1, i.e., the achievable minimum NDT for the worst-case user demand when the centralized cache placement as in \cite{Alilimits} is adopted.

 We consider the normalized cache size $\mu\in\left\{0,\frac{1}{K},\frac{2}{K},\ldots,1\right\}$ such that the accumulated cache size $K\mu$ is an integer\footnote{For the general case where $K\mu$ is not an integer, the memory sharing technique shall be used.}. Each file $W_{\ell}$, for $\ell\in[L]$, is equally partitioned into $\binom{K}{K\mu}$ disjoint subfiles as
 \begin{equation}
   W_{\ell}=\{W_{\ell, {\mathcal K}}: |{\mathcal K}|=K\mu, {\mathcal K}\subseteq[K]\},
 \end{equation}
 and each subfile $W_{\ell,{\mathcal K}}$ is cached at user subset $\mathcal K$. The lengths of these subfiles are the same and given by
 \begin{equation}
   |W_{\ell, {\mathcal K}}|=\frac{F}{\binom{K}{K\mu}}\quad  \text{bits}.
 \end{equation}
For the worst-case user demand where each user requests a distinct file, for any ${\mathcal A}\subseteq[K]$ with size $s=K\mu+1$, we can generate multicast messages as \eqref{generate_coded_message}. The lengths of these coded messages are the same and given by
\begin{equation}
  a_{\mathcal A}=\frac{1-\mu}{\binom{K-1}{s-1}}, \quad \forall {\mathcal A}\in\Psi.
\end{equation}
Due to this property, by using the constraint \eqref{P1_time}, each $d_{\mathcal A}$ of these coded messages in the DoF tuple should be the same for all ${\mathcal A}\in \Psi$. According to Theorem \ref{theorem2}, we can obtain the per-group DoF $d$ as
\begin{align}
  &d=   \nonumber \\
  &\begin{cases}
           \frac{M}{\binom{K}{s}}, & \frac{M}{N}\in\left(0,\frac{s\binom{K}{s}}{1+(s-1)\binom{K}{s}}\right]\\
           \max\left\{\frac{M}{K\binom{K-1}{s-1}}, \frac{sN}{1+(s-1)\binom{K}{s}}\right\}, & \frac{M}{N}\in\left(\frac{s\binom{K}{s}}{1+(s-1)\binom{K}{s}},K\right] \\
           \frac{N}{\binom{K-1}{s-1}},& \frac{M}{N}\in(K,\infty) \\
           \end{cases}.
\end{align}
The corresponding NDT can be calculated as
\begin{equation}
  \tau_{\text{worst}}=\frac{1-\mu}{\binom{K-1}{s-1}}\cdot\frac{1}{d},
\end{equation}
which completes the proof of Corollary 1.

\section{Proof of Corollary 2}\label{proof_of_corollary_2}
In this section, we present the proof of Corollary 2, i.e., the achievable minimum NDT for the worst-case user demand when the decentralized cache placement as in \cite{Alidecentralized} is adopted and the file length is large.

Considering the worst-case user demand, the generation of coded messages can be done as \eqref{generate_coded_message}.
 By using the law of large numbers, the length of coded message $W_{\mathcal A}^{\oplus}$ is approximately given by
\begin{equation}
  a_{\mathcal A}=\mu^{s-1}(1-\mu)^{K-s+1},
\end{equation}
where $s$ denotes the size of multicast group $\mathcal A$. In this case, the size $s$ can vary from $1$ to $K$. Since the lengths of coded messages intended for equal-size multicast groups are the same, by using the constraint \eqref{P1_time}, the corresponding DoFs of these coded messages in the DoF tuple should also be the same, i.e.,
\begin{equation}\label{equal_d}
  d_{\mathcal A}=d_s, \quad \forall|\mathcal A|=s.
\end{equation}
\subsection{$\frac{M}{N}\in\left (0,1 \right]$}
In this case, transmitting these coded messages one by one can also achieve the optimal DoF region (see Remark \ref{tdd_optimal}). Thus, the achievable minimum NDT for all the coded messages can be calculated as
\begin{equation}
  \tau_{\text{worst}}=\sum_{s=1}^{K}\binom{K}{s}\cdot \mu^{s-1}(1-\mu)^{K-s+1}\cdot \frac{1}{M}.
\end{equation}

\subsection{$\frac{M}{N}\in\left (K,\infty \right)$}
According to the proof of Theorem \ref{theorem2}, all the coded messages can be sent simultaneously at this antenna configuration. Each user can receive its desired coded messages at DoF of $N$. Thus, the achievable minimum NDT can be calculated as
\begin{align}
  \tau_{\text{worst}}&=\sum_{s=1}^{K}\binom{K-1}{s-1}\cdot \mu^{s-1}(1-\mu)^{K-s+1}\cdot \frac{1}{N} \\
        &=\frac{1}{N}\sum_{q=0}^{K-1}\binom{K-1}{q}\mu^{q}(1-\mu)^{K-q}  \\
        &=\frac{1-\mu}{N}(1-u+u)^{K-1}\\
        &=\frac{1-\mu}{N}.
\end{align}

\subsection{$\frac{M}{N}\in\left (1,K \right]$}
Although we cannot obtain the closed-form NDT in this case, by using Theorem \ref{theorem2} and condition \eqref{equal_d}, the region ${\mathcal D}_2$ can be simplified as \eqref{simplified_d}. It can be seen that the total number of inequalities in $\tilde{{\mathcal D}}_2$ is reduced from $2^K+1$ to $K+1$, which is helpful to solve the problem ${\mathcal P}_1$ more efficiently.

So far, we complete the proof of Corollary 2.

\end{appendices}

\bibliographystyle{IEEEtran}
\bibliography{IEEEabrv,reference}

\end{document}